\documentclass[journal]{IEEEtran}
\usepackage{amsmath,amsfonts}
\usepackage{algorithmic}
\usepackage{algorithm}
\usepackage{array}
\usepackage[caption=false,font=normalsize,labelfont=sf,textfont=sf]{subfig}
\usepackage{textcomp}
\usepackage{stfloats}
\usepackage{url}
\usepackage{verbatim}
\usepackage{graphicx}
\usepackage{cite}
\graphicspath{{./figures/}}
\begin{document}

\title{Collaborative Multi-BS Power Management for Dense Radio Access Network using Deep Reinforcement Learning}

\author{Yuchao~Chang,~Wen~Chen,~\IEEEmembership{Senior~Member,~IEEE},~Jun~Li,~\IEEEmembership{Senior~Member,~IEEE},
~Jianpo~Liu,~Haoran~Wei,~\IEEEmembership{Member,~IEEE},~Zhendong~Wang,~and~Naofal~Al-Dhahir,~\IEEEmembership{Fellow,~IEEE}
        % <-this % stops a space
\thanks{Y. Chang and W. Chen are with the Department of Electronic Engineering, Shanghai Jiao Tong University, Shanghai 200240, China (e-mail: yuchaoc@mail.ustc.edu.cn; wenchen@sjtu.edu.cn).}% <-this % stops a space
\thanks{J. Li is with the School of Electronic and Optical Engineering, Nanjing University of Science Technology, Nanjing 210094, China (email: jun.li@njust.edu.cn).}
\thanks{J. Liu is with Science and Technology on Microsystem Laboratory, Shanghai Institute of Microsystem and Information Technology, Chinese Academy of Sciences, Shanghai 201800, China (email: liujp@mail.sim.ac.cn).}
\thanks{H. Wei , Z. Wang, and N. Al-Dhahir are with Department of Electrical and Computer Engineering, The University of Texas at Dallas, Richardson, TX 75080, USA (email: haoran.wei@utdallas.edu; zhendong.wang@alumni.uml.edu; aldhahir@utdallas.edu).}% <-this % stops a space
\thanks{Manuscript received November 02, 2022; revised February 28, 2023 and March 29, 2023.}
\thanks{(\emph{Corresponding author: Wen Chen.})}
}
\markboth{Journal of \LaTeX\ Class Files,~Vol.~14, No.~8, August~2021}%
{Shell \MakeLowercase{\textit{et al.}}: Collaborative Multi-BS Power Management for Dense Radio Access Network using Deep Reinforcement Learning}


\maketitle

\begin{abstract}
Network energy efficiency is a main pillar in the design and operation of wireless communication systems. In this paper, we investigate a dense radio access network (dense-RAN) capable of radiated power management at the base station (BS). Aiming to improve the long-term network energy efficiency, an optimization problem is formulated by collaboratively managing multi-BSs' radiated power levels with constraints on the users' traffic volume and achievable rate. Considering stochastic traffic arrivals at the users and time-varying network interference, we first formulate the problem as a Markov decision process (MDP) and then develop a novel deep reinforcement learning (DRL) framework based on the cloud-RAN operation scheme. To tackle the trade-off between complexity and performance, the overall optimization of multi-BSs' energy efficiency with the multiplicative complexity constraint is modeled to achieve near-optimal performance by using a deep Q-network (DQN). In DQN, each BS first maximizes its individual energy efficiency, and then cooperates with other BSs to maximize the overall multi-BSs' energy efficiency. Simulation results demonstrate that the proposed algorithm can converge faster and enjoy a network energy efficiency improvement by 5\% and 10\% compared with the benchmarks of the Q-learning and sleep schemes, respectively.
\end{abstract}


\section{Introduction}
\IEEEPARstart{W}{ireless} communications technology has progressed from the first generation (1G) to the latest fifth generation (5G) and beyond to meet the increasing and diversified traffic requirements \cite{HeD2019COMST, XuY2019JSAC}. The anticipated more than 1000-fold increasing requirements of wireless traffic and global recognition of green communications pose significant challenges for 5G and beyond wireless communication design \cite{SaxenaN2016JSAC}. There are three basic means to improve wireless communication performance: (1) network densification, (2) more spectrum, and (3) increasing the spectral efficiency \cite{marzetta_larsson_yang_ngo_2016}. Therefore, future wireless communication standards are likely to deploy access points with higher density, to use new spectral bands, and to maximize the spectral efficiency \cite{LiaoW2019TVT, SongH2011TTHZ}. Since the power consumption of a single 5G base station (BS) is more than 1.5 times that of a 4G BS, and the number of required 5G BSs is more than 4 times that of 4G networks per coverage area \cite{mao2021ai}, the power consumption of 5G networks per coverage area is at least 6 times that of 4G networks. Therefore, reducing the power consumption of 5G networks is a critical task, which implies that green communication with high energy-efficiency is a key design goal \cite{HasanZ2011SURV, s17071665, WangKL2020TGCN}. Meanwhile, different from the conventional cell-centric wireless communications, 5G and beyond networks are more likely to be user-centric systems due to diversified user service demands, implying that future wireless communications need to be more intelligent for effective management \cite{NguyenDC2021COMST}. Hence, 5G and beyond wireless communications embraces two key themes: \emph{green} and \emph{intelligence}.
\par
Currently, the global carbon emissions have continued to rise for years, resulting in more harsh global weather and severe air pollution. Meanwhile, unlike the third generation (3G) and the fourth generation (4G) wireless communications, 5G and beyond wireless communications are going to be faced with huge amounts of mobile traffic that is generated by tens of billions mobile terminals \cite{XuY2019JSAC, YChangACCESS2019}. Mao \emph{et al} in \cite{mao2021ai} pointed out that the information and communications technology (ICT) industry accounts for nearly 20\% of the total electricity consumption and keeps an annual growth rate of more than 6\%. That paper also presented main considerations for green communications and research on artificial intelligence-based (AI-based) green communications. In 2021, ICT's share of global greenhouse gas (GHG) emissions was estimated to be 1.8-2.8\% of the global GHG emissions \cite{ICT2021The}. Many researchers have shown the possibility of improving energy efficiency from different perspectives. The authors in \cite{mahapatra2015energy} conducted an extensive survey of energy efficiency and discussed many tradeoff techniques in green communications. The research in \cite{zhang2016fundamental} analyzed four basic network tradeoffs: spectrum efficiency versus energy efficiency, deployment efficiency versus energy efficiency, delay versus power, and bandwidth versus power. The survey on green mobile networking in \cite{ismail2014survey} discussed detailed modeling methods for power consumption and energy efficiency. In \cite{TACT2014_RLi}, Li \emph{et al} clearly pointed out that over 80\% of the power consumption took place in the radio access networks (RAN), implying that the sleeping scheme is a very useful strategy for improving network energy efficiency \cite{BBDai2016JSAC, RTao2019TWC, MQinTWC2020, IEEE.Network.2021.Shinkuma}. To minimize the network power consumption, Dai \emph{et al} developed an energy-efficient design of downlink C-RAN based on the low-power sleeping mode while considering user rate constraints \cite{BBDai2016JSAC}. A novel sleeping mechanism for heterogeneous network (HetNets) was proposed to decrease energy consumption of multiple BSs by considering traffic dynamics \cite{RTao2019TWC}. Furthermore, some resource allocation schemes have been proposed to maximize network energy efficiency by exploring the BS switching operations. Qin \emph{et al.} in \cite{MQinTWC2020} proposed a resource-on-demand energy scheduling strategy to make effective utilization of harvested energy. A multi-objective auction-based switching-off scheme in heterogeneous networks was developed to improve energy and cost savings \cite{TVT2016Bousia}. In \cite{TVT.2017.2719404}, the work enhanced the energy efficiency of HetNets by switching off a part of small cells in HetNets. The works in \cite{matthiesen2020globally, dong2020deep} utilize a neural network (NN) to optimize the transmit power aiming at minimizing the network energy consumption. Multiple access technologies, for example rate splitting multiple access (RSMA) \cite{TCOMM.2021.3091133} and non-orthogonal multiple access (NOMA) \cite{ding2017survey}, also play an important role in enhancing power management and has achieved positive results in RAN. Although the above contributions have improved energy efficiency of wireless communication, they fail to consider accurate downlink power management.
\par
In 5G and beyond communication systems, the network scale and complexity have increased substantially, which makes traditional network optimization strategies inefficient and ineffective. Thanks to the abilities of AI technology in tackling large-scale and complex tasks, it can significantly improve the efficiency of network optimization \cite{shi2020communication, BennisM2013TWC, ChangY2018LCOMM, feriani2021single}. Hence, similar to \emph{green},  {\emph{intelligence}} is also a critical characteristic of 5G and beyond wireless communications. Being a key approach in machine learning, reinforcement learning (RL) is an environment-based strategy and is realized through continuous interaction in discrete time steps between the agent and environment. To maximize the accumulative reward, RL continuously learns how to take actions automatically by interacting with the environment \cite{Book_RL}. The basic elements of RL are a finite set of states, a finite set of actions, and the reward function. In each episode, the agent identifies the environment state and then takes action based on the optimal policy. At the end of the episode, the agent evaluates the action based on the reward function. RL has been applied to a wide range of engineering applications such as healthcare and autonomous control \cite{MMahmud2018TNNLS, liu2013, Sustainability2022, BKiumarsi2018TNNLS}. For example, the actor-critic technology has been adopted to conserve network energy in \cite{TACT2014_RLi,wei2017user}. Wang \emph{et al.} adopted deep Q-network (DQN) to identify the optimal policy in channel access, where the dynamic channel access problem was modeled as a partially observable Markov decision process (POMDP) \cite{SWang2018TCCN}. To explore a representative user's computation offloading to multi-BSs, Chen \emph{et al.} adopted DQN to determine the optimal offloading strategy \cite{XChen2019IoT}.
In a MIMO-based UAV network, the authors in \cite{HHuang2020TVT} investigated DQN to maximize its coverage efficiency calculated based on the received signal strength. As a well-established artificial neural network (ANN) architecture, the back propagation neural network (BPNN) is ideal as a DQN core and commonly used learning algorithm for complex function approximation \cite{AlSammarraie2018ICSCEE}. For example, the authors in \cite{WangY2018TII} proposed a novel adaptive learning hybrid model for precise solar intensity forecasting by using BPNN. A general BPNN is designed to consist of one input layer, multiple hidden layers, and one output layer. Considering the complexity of future wireless communication networks, the reinforcement learning techniques have great potential to improve the communication performance by using both the continuous environmental interaction and the NN's function approximation.
\par
In \cite{JMurdock2011GLOCOMW}, energy efficiency (EE) was proposed to evaluate the network energy efficiency of a wireless communication link. The energy efficiency is defined as the ratio of data rate to consumed power. Hence, to improve the network energy efficiency, this paper proposes a novel precise downlink power model by using deep reinforcement learning (DRL) based on BPNN to control power intelligently. The main contributions of this paper are summarized as follows:
\begin{itemize}
\item In dense-RAN, we propose a novel adaptive multiple base stations (multi-BSs) downlink power management model to improve network energy efficiency. In the studied model, based on the unpredictable differentiated traffic demands of users, the central processor simultaneously optimizes the downlink power of multi-BSs while guaranteeing the throughput requirement.
\item To tackle the power management model, we design a DRL framework, where the central processor serves as an agent. The downlink power levels and network energy efficiency are the action space and reward function, respectively. %DQN is adopted to optimize the multi-BSs' power management to maximize the cumulative reward.
\item Considering the trade-off between complexity and performance, the network energy efficiency optimization of the DRL framework is modeled to achieve near-optimal precise power management, where the cumulative energy efficiency is maximized by using a DQN with multiplicative complexity. Moreover, BPNN is used to tune the considered DQN parameter.
\item Simulation results show that our proposed algorithm enjoys a network energy efficiency improvement of 5\% and 10\% compared with that of the Q-learning and sleep schemes, respectively.
\end{itemize}
\par
This paper is structured as follows: Section \ref{section:SysMdl} describes the system model, where the radio communication model and problem formulation are described, respectively.
Section \ref{section:AFBRL} presents the deep reinforcement learning framework, in which we model the precise downlink power management problem as MDP, and adopt DQN based on BPNN to identify the optimal policy for precisely managing downlink power.
Section \ref{section:NumericalResults} demonstrates the superiority of the proposed algorithm in terms of both communication performance and computational complexity. Section \ref{section:conclusion} concludes our work. Moreover, for ease of reference, \textbf{Table} \ref{tab:No01} lists the main acronyms.
\begin{table}[!t]
\caption{Summary of main acronyms}
\centering
\label{tab:No01}
\begin{tabular}{|c||c|}
\hline
\textbf{Acronym} & \textbf{Meaning} \\
\hline
  1/2/3/4/5G & First/second/third/fourth/fifth generation \\
  \hline
  ANN & Artificial neural network \\
  \hline
  BPNN & Back propagation neural network   \\
  \hline
  BS & Base station \\
  \hline
  EE & Energy efficiency \\
  \hline
  DRL & Deep reinforcement learning  \\
  \hline
  DQN & Deep Q-network \\
  \hline
  dense-RAN & Dense radio access network \\
  \hline
  GHG & Global greenhouse gas  \\
  \hline
  MDP &  Markov decision process \\
  \hline
  POMDP & Partially observable markov decision process \\
  \hline
  RSRP & Reference Signal Receiving Power   \\
  \hline
  SINR & Signal-to-interference-plus-noise-ratio \\
  \hline
  UMa & Urban Macro \\
\hline
\end{tabular}
\end{table}
\begin{table}[!t]
\caption{Summary of main notation}
\centering
\label{tab:No02}
\begin{tabular}{|c||c|}
\hline
\textbf{Notation} & \textbf{Description} \\
\hline
  $\varphi _{[b^{'},t]}$ & Active or sleep status of BS $b^{'}$ \\
  \hline
  ${\gamma _{\left[ {bu,t} \right]}}$ & SINR of user $u$ \\
  \hline
  $\lambda$ & Discount factor \\
  \hline
  $\pi \left( {s,a} \right)$ & Transition probability \\
  \hline
  $h_{[b,t]}$ & Radiating signal from BS $b$ at time $t$ \\
  \hline
  $l_b$, $l_{[u,t]}$  & Coordinate of BS $b$ and user $u$ \\
  \hline
  $o_b$, $o_{[u,t]}$  & Height of BS $b$ and user $u$ \\
  \hline
  $y_{[u,t]}$ & Y-coordinate for user u at time $t$ \\
  \hline
  ${\mathcal A}$ & Action set in RL  \\
  \hline
  ${\mathcal B}$  & Set of BSs  \\
  \hline
  $C_ {[bu,t]}$ & Data rate from BS $b$ to user $u$ \\
  \hline
  $\overline C _t^\Sigma$ & Cumulative average throughput \\
  \hline
  ${\mathcal{D}}$  & Replay memory \\
  \hline
  ${\mathcal D}_m$  & Mini-batch memory  \\
  \hline
  $\operatorname{D}\!\left( {{l_b},{l_{[u,t]}}} \right)$  & Distance between BS $b$ and user $u$ \\
  \hline
  ${G_{\left[ {bu,t} \right]}}$ &  Total discounted rewards in RL \\
  \hline
  $H_{[bu,t]}$  &  Channel gain from BS $b$ to user $u$ \\
  \hline
  $\operatorname{L}\left( {{\mathbf W} _{\text{p}}}, {{\mathbf W} _{\text{o}}} \right)$ & Loss function \\
  \hline
  ${\overline N _T^{*,\Sigma }}$ & Overall average number of iterations \\
  \hline
  ${\overline N _t^{*,\Sigma }}$ & Cumulative average number of iterations \\
  \hline
  ${\mathcal P}$  &  Power levels' set   \\
  \hline
  $P_{[b,t]}$  &  Power level of BS $b$   \\
  \hline
  $\overline P _t^\Sigma$ & Cumulative average power \\
  \hline
  $\Delta \overline P _t^{S,\Sigma }$ &  Cumulative average RSRP decline  \\
  \hline
  $\Delta \overline P _t^{I,\Sigma }$ &  Cumulative average interference decline  \\
  \hline
  $Q_{\left[ {bu,t} \right]}^\pi \left( {s,a} \right)$ & State-action-value function  \\
  \hline
  $R_t$ & Network energy efficiency  \\
  \hline
  $\overline R _T^{\Sigma}$ & Overall average EE \\
  \hline
  $\overline R _t^{\Sigma}$ & Cumulative average EE \\
  \hline
  ${\mathcal S}$ & State set in RL  \\
  \hline
  ${\mathcal T}$  & Set of time-steps  \\
  \hline
  ${\mathcal U}$  & Set of users  \\
  \hline
  ${V_{[bu,t]}}$ & Traffic volume from BS $b$ to $u$  \\
  \hline
  ${\mathbf W} _{\text{o}}$ & DQN parameters of target \\
  \hline
  ${\mathbf W} _{\text{P}}$ & DQN parameters of predicted \\
  \hline
  $Y_{[u,t]}$ & Received signal of user $u$ in time-step $t$ \\
  \hline
  $ {\overline Z _T^{\Sigma }}$ & Overall average success ratio \\
  \hline
  ${\overline Z _t^{\Sigma }}$ & Cumulative average success ratio \\
\hline
\end{tabular}
\end{table}
\begin{figure}[!t]
  \centering
  \includegraphics[width=2.5in]{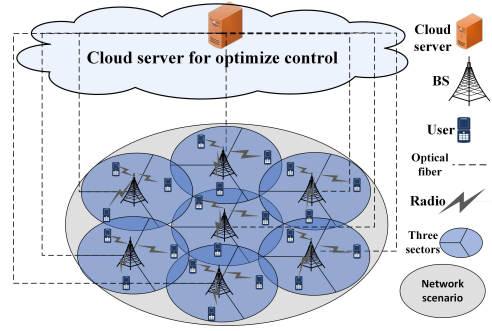}\\
  \caption{The DRL-RAN architecture deploying BSs and users.}\label{Fig_01_RRU_TU_Dis}
\end{figure}
\section{SYSTEM MODEL}\label{section:SysMdl}
The studied system model is analyzed in a dense-RAN urban scenario, where downlink power management for all BSs is explored. Meanwhile, to reduce the back-lobe interference of the BS sector antenna, each BS is designed to include three sectors. In particular, each sector has its own antenna and serves different users and is independently laid out in its related BS \cite{SystemModel_EuCAP_RYamaguchi}. The studied dense-RAN architecture considers a geographical area, where the set of BSs, denoted by ${\mathcal B} = \left\{ {1, \ldots ,b, \ldots ,B} \right\}$, are pre-deployed by the operator and a group of randomly deployed terrestrial users, denoted by ${\mathcal U} = \left\{ {1, \ldots ,u, \ldots ,U} \right\}$, generate unpredictable differentiated service demands over time. The dense-RAN scenario is illustrated in \textbf{Fig. }\ref{Fig_01_RRU_TU_Dis}. Specifically, the system model includes a cloud server for optimization control and the network scenario for deploying BSs and users. The cloud server communicates with BSs through the optical fiber, while the BS communicates with the user through the radio channel. For example, the coordinates of the BS $b$ and the user $u$ in time-step $t$ are given by
\begin{equation}\label{SysMdl_RRU_TU_Location}
    \left\{ \begin{gathered}
    {l_b} = \left( {{x_b},{y_b},{o_b}} \right), \qquad \qquad \qquad \ \ \forall b \in {\mathcal B}; \hfill \\
    {l_{[u,t]}} = \left( {{x_{[u,t]}},{y_{[u,t]}},{o_{[u,t]}}} \right), \qquad \forall u \in {\mathcal U}; \hfill \\
    \end{gathered}  \right.
\end{equation}
where $x_b$, $y_b$, $x_{[u,t]}$, and $y_{[u,t]}$ are plane coordinates, while $o_b$ and $o_{\left[ {u,t} \right]}$ are the heights. Moreover, the distance between BS $b$ and user $u$ in time-step $t$ is calculated using the relation
\begin{equation}\label{SysMdl_RRU_TU_Distance}
    \begin{gathered}
      \operatorname{D}\!\left( {{l_b},{l_{[u,t]}}} \right)=\hfill \\
      \sqrt {{{\left( {{x_b} - {x_{[u,t]}}} \right)}^2}+{{\left( {{y_b} - {y_{[u,t]}}} \right)}^2}\!+\!{{\left( {{o_b} - {o_{[u,t]}}} \right)}^2}},  \hfill \\
    \end{gathered}\!\!
\end{equation}
where $\operatorname{D}\left( {{l_b},{l_{[u,t]}}} \right)$ is a smooth function related to location change of user $u$ over time.
\par
In long-term energy-efficient wireless networks, the user traffic arrives randomly and network interference fluctuates over time. Hence, we adopt the time-step structure in the dense-RAN system, where the operating period is discretized into the time-step space ${\mathcal T} = \left\{ {0,1, \ldots ,T - 1} \right\}$, where every time-step is regulated by 1 msec. In orthogonal frequency division multiple access (OFDMA), the different frequency sub-carriers are orthogonal to each other. Hence, we optimize transmit power allocations between BSs at the same frequency sub-carriers. Power management is more conducive to improving network energy efficiency by interference coordination of the same frequency sub-carriers. Hence, we configure an OFDMA-based network scenario for the dense-RAN, where transmit power allocations at the same frequency sub-carriers are optimized. Considering that the network scheduling optimization is out of our research scope, the associations between BSs and users are known in advance based on the maximal Reference Signal Receiving Power (RSRP) access. This indicates that one user can choose the BS with the maximal RSRP as its association BS when it has many candidate BSs. For ease of reference, \textbf{Table} \ref{tab:No02} lists the main notations of the system model.
\subsection{Radio Communication Model}\label{section:SysMdl_RCM}
Considering our research on power management, we simplify the channel propagation model and reduce the estimation overhead by using widebeams \cite{Lee2019SPAWC}. Thus, we adopt the link budget calculation to explore the downlink power management optimization problem in the studied dense-RAN \cite{AidaLopezMWC2019}. The link budget is a summary of the transmit power along with all the link's gains and losses, which enables the strength of the received signal at the user to be calculated. The strength of the received signal is measured by RSRP \cite{RoseHu2013Book}. It is possible to estimate whether RSRP at the user is sufficient, too high, or too low, implying that the corrective transmit power at the BS is managed for ensuring satisfactory system operation. When it is larger than the required RSRP for the user, the high transmit power at the BS can be reduced to minimize the power cost while still meeting the communication demand of its associated user.
\par
Assume that BS $b$ is serving user $u$ and any other BS $b^{'} ( b^{'} \in {\mathcal B}, b^{'}\neq b )$ is the interference BS. Then, the received signal of user $u$ in time-step $t$, including the serving RSRP and interference RSRP, can be expressed as follows
\begin{equation}\label{SysMdl_LinkBudge_dB}
    \begin{aligned}%{gathered}
      {Y_{[u,t]}} &= \underbrace {{h_{[b,t]}}{s_{[u,t]}}}_{{\text{desired signal}}} + \underbrace {\sum\limits_{{b^{'}} \in {\mathcal B}\backslash b} {{h_{{[b^{'},t]}}}{s_{[u^{'},t]}}} }_{{\text{inter-user interference signal}}} + \underbrace {{n_0}}_{{\text{noise signal}}} \hfill \\
       &= \left ( {P_{[b,t]}}{H_{[bu,t]}} \right ) {s_{[u,t]}} \hfill \\
       & \qquad + \sum\limits_{{b^{'}} \in {\mathcal B} \backslash b} {\left ( {P_{{[b^{'},t]}}}{H_{[{b^{'}}u,t]}} \right ) {s_{[{u^{'}},t]}}}  + {n_0},  \hfill \\
    \end{aligned}%{gathered},
\end{equation}
where $h_{[b,t]}={P_{[b,t]}}{H_{[bu,t]}}$ and $h_{[b^{'},t]}={P_{[b^{'},t]}}{H_{[b^{'}u,t]}}$. ${P_{[b,t]}}$ and ${P_{[b^{'},t]}}$ $\left ( P_{[b,t]}, P_{[b^{'},t]} \in {\mathcal P} \right )$ are, respectively, the downlink power levels of BS $b$ and $b^{'}$, and ${\mathcal P} = \left\{ {{P_0}, \ldots, {P_{\kappa-1} }} \right\}$ is the set of available downlink power levels for all BSs. In addition, $s_{[u,t]}$ and $s_{[{u^{'}},t]}$ are the received signals for user $u$ and $u^{'}$, respectively, and $n_0$ is the Gaussian white noise signal at the user.
\par
Moreover, ${H_{[bu,t]}}$ is the gain that accounts for the total transmitter gain, path loss, and total receiver gain from BS $b$ to user $u$ in time-step $t$, while ${H_{[b^{'}u,t]}}$ is defined similar to ${H_{[bu,t]}}$ and is the channel gain from BS $b^{'}$ to user $u$. In particular, based on the UMa (Urban Macro) propagation model in the 3GPP TR 38.901 standard \cite{3gpp2018study}, the channel gain from BS $b$ to user $u$ in time-step $t$ is given by \cite{NurEsa2020ICIEE5GNR}
\begin{equation}\label{SysMdl_Channel_Gain}
    \begin{aligned} %{gathered}
      {H_{\left[ {bu,t} \right]}} &= {H_{TX}} \times {PL_{[bu,t]}} \times {H_{RX}} \hfill \\
      & = {H_{TX}} \times \left( {\frac{c}{{4\pi  \times {f_{c}} \times \operatorname{D}\left( {{l_b},{l_{\left[ {u,t} \right]}}} \right)}}} \right)  \times {H_{RX}}, \hfill \\
    \end{aligned} %{gathered} ,
\end{equation}
where $H_{TX}$ and $H_{RX}$ are, respectively, the signal gain constants at the transmitter and receiver. $PL_ {[bu,t]}$ is the path loss from BS $b$ to user $u$, c is the speed of light, and $f_{c}$ is the center frequency.
\par
Accordingly, based on the definition in (\ref{SysMdl_LinkBudge_dB}), the signal-to-interference-plus-noise-ratio (SINR) for user $u$ in time-step $t$ is given by
%\begin{small}
\begin{equation}\label{SysMdl_SINR_TUn}
    \begin{aligned} %{array}{l}
    {\gamma _{\left[ {bu,t} \right]}} &= \frac{{{h_{[bu,t]}}}}{{\sum\limits_{{b^{'}} \in {\mathcal B}\backslash b} {\left ( {\varphi _{{[b^{'},t]}}}{h_{[b^{'}u,t]}} \right ) }  + {\sigma ^2}}}\\
     &= \frac{{{P_{\left[ {b,t} \right]}}{H_{\left[ {bu,t} \right]}}}}{{\sum\limits_{{b^{'}} \in {\mathcal B}\backslash b} {\left ( {\varphi _{{[b^{'},t]}}}{P_{\left[ {{b^{'}},t} \right]}}{H_{\left[ {{b^{'}}u,t} \right]}} \right ) }  + {\sigma ^2}}},
    \end{aligned} %{array}
\end{equation}
where $\sigma ^2$ is the noise power. In time-step $t$, $\varphi _{[b^{'},t]} = 1$ when BS $b^{'}$ is active for serving a user, and $\varphi _{[b^{'},t]} = 0$ when BS $b^{'}$ is sleep for serving no user. Then, the downlink data rate from BS $b$ to user $u$ in time-step $t$ is given by
\begin{equation}\label{SysMdl_CM_DR}
    {C_{[bu,t]}} = W{\log _2}\left( {1 + {\gamma _{[bu,t]}}} \right),
\end{equation}
where $W$ is the channel bandwidth. We assume that the downlink data rate of user $u$ is $C_{bu}^{\max }$ when BS $b$ transmits data with the standard downlink power level $P_{\text{max}}$. The power decreases of BS $b$ and the downlink data rate variation in time-step $t$ are, respectively, expressed as follows
\begin{align}
\Delta {P_{[b,t]}} &= {\varphi  _{[b,t]}}\left( {{P_{\max }} - {P_{[b,t]}}} \right), \label{SysMdl_Energy_Conservation} \\
\Delta {C_{[bu,t]}} &= {\varphi  _{[b,t]}}\left( {{C_{[b,t]}^{\max }} - {C_{[b,t]}}} \right). \label{SysMdl_TGOLos}
\end{align}
where $\varphi  _{[b,t]}$ is defined similar to $\varphi  _{[b^{'},t]}$ with BS $b$.
\par
By dividing the downlink data rate in (\ref{SysMdl_CM_DR}) by the downlink power, the network energy efficiency is given in (\ref{SysMdl_CM_PowerEfficiency}) in units of bits per seconds per dBW (Mbps/dBW) \cite{JMurdock2011GLOCOMW}.
\begin{equation}\label{SysMdl_CM_PowerEfficiency}
  { R _t} = \frac{1}{{\sum\limits_{b = 1}^B {{\varphi _{\left[ {b,t} \right]}}} }}\sum\limits_{b = 1}^B {{R_{\left[ {bu,t} \right]}}},
\end{equation}
where
\begin{equation}\label{SysMdl_CM_PowerEfficiency_1}
  {R_{\left[ {bu,t} \right]}} =  {\frac{{{C_{\left[ {bu,t} \right]}}}}{{{P_{\left[ {b,t} \right]}}}}},
\end{equation}
and the power unit translation between "dBW" and "W" is denoted as
\begin{equation}\label{SysMdl_CM_PowerTranslation}
    P\left[ {{\rm{in \  dBW}}} \right] = 10 \cdot {\log _{10}}\left( {\frac{{P\left[ {{\rm{in \  W}}} \right]}}{{1\left[ {{\rm{in \  W}}} \right]}}} \right).
\end{equation}
\par
Besides, when user $u$ requests a traffic volume ${V_{[bu,t]}}$ from its associated BS $b$ in time-step $t$, the number of required time-steps starting from time-step $t$, denoted by ${T_{[bu,t]}}$, satisfies the following constraint
\begin{equation}\label{SysMdl_CM_NTSofEPO}
    {V_{[bu,t]}} = \sum\limits_{{k =0}}^{{T_{[bu,t]}}-1} {C_{[bu,t+k]}},
\end{equation}
where  $T_{[bu,t]} \leq T$, and it is similar to the relation between any other BS $b^{'}$ and its corresponding associated user.
\subsection{Problem Formulation}\label{section:SysMdl_ProblemFormation}
In dense-RAN, based on constraints of the traffic volume and downlink data rate variation, the long-term energy efficiency maximization problem is formulated as follows
\begin{equation}\label{SysMdl_ProblemFormation}
    \begin{gathered}
        \mathop {\operatorname{maximize} } \sum\limits_t {{R_t}} ; \hfill \\
        s.t. \quad \sum\limits_{{k =0}}^{{T_{[bu,t]}}-1} {C_{[bu,t + k]}} = {V_{[bu,t]}}; \hfill \\
        \qquad \ \ \sum\limits_{b=1}^{B} {{\Delta C _{[ {bu,t} ]}}}  \geq 0 ; \hfill \\
        \qquad \ \ {P_{[b,t]}} \in {\mathcal P}; \hfill \\
        \qquad \ \
        b \in {\mathcal B}, \
        u \in {\mathcal U}, \
        t \in {\mathcal T}. \hfill \\
    \end{gathered}
\end{equation}
where the power level ${P_{[b,t]}}$ for any BS $b (b \in {\mathcal B})$ in time-step $t$ is the optimization variable, the network energy efficiency $\sum\limits_t {{R_t}}$ is the optimization objective, and other parameters, for example the association between BS and user $\varphi _{[b^{'},t]}$ and the power set ${\mathcal P}$, are the scenario setting.
The first constraint is the traffic volume determined by the traffic volume demand for any active user $u (\forall u \in {\mathcal U})$ and its serving BS $b$. The second constraint shows the network throughput limit, which is calculated by accumulating downlink data rate variations for all BSs in any time-step $t (\forall t \in {\mathcal T})$. The third constraint illustrates that the available downlink power levels are identical for all BSs. By combining formulas (\ref{SysMdl_SINR_TUn}) and (\ref{SysMdl_CM_DR}), the downlink data rate in the first constraint of the optimization problem (\ref{SysMdl_ProblemFormation}) is computed as follows
\begin{equation}\label{SysMdl_ProblemFormation_Rate}
    \begin{gathered}
      C_{[bu,t + k]} = \hfill \\
      W{\log _2}\!\!\left(\!\!{1\!+\!{\frac{{{P_{\left[ {b,t + k} \right]}}{H_{\left[ {bu,t + k} \right]}}}}{{\sum\limits_{{b^{'}} \in {\mathcal B}\backslash b}\!\!\!{\left ( {\varphi _{{[b^{'},t + k]}}}{P_{\left[ {{b^{'}},t + k} \right]}}{H_{\left[ {{b^{'}}u,t + k} \right]}} \right ) }\!+\!{\sigma ^2}}}}} \right)\!\!.\hfill \\
    \end{gathered}
\end{equation}
\section{Deep Reinforcement Learning Framework}\label{section:AFBRL}
Wireless random traffic has made large-scale and complex network optimization techniques move from the rule-based to AI-based. The RL's learning process is realized through discrete time steps, implying that RL is a better alternative for the dense-RAN scenario. As an example RL model, DQN has been introduced to predict the Q-value for tackling the continuous and large state space and action space by using deep neural networks. In \cite{DBLP2018}, DQN was firstly proposed to train an agent to learn a policy from its observations by Google DeepMind based on an ANN, which overcomes the limitation of the traditional look-up table approach in Q-learning. The actual solution of this optimization problem is to obtain the near-optimal collaborative multi-BS power management while meeting the network performance requirements \cite{JSAC2016Mai}. RL is a continuous learning algorithm that optimizes control through continuous observations, which can deal with the changing state of complex systems in real time and realize an optimized control scheme. Moreover, we learn that the traffic volume possesses Markov process characteristics from (\ref{SysMdl_CM_NTSofEPO}). Hence, it is feasible to formulate the optimization problem in (\ref{SysMdl_ProblemFormation}) as an MDP problem. Our goal is to maximize the long-term network energy efficiency with user traffic volume demand while guaranteeing no network throughput loss.
\subsection{Reinforcement Learning in a Nutshell}\label{section:AFBRL_RLN}
The system state is designed with the residue transmitted traffic volume and the users' RSRP. In particular, the state space is defined as follows
\begin{equation}\label{AFBRL_State_Space}
    {\mathcal S} = \left\{ {{S_t}\left| \begin{gathered}
      {S_t} = \left\{ {{V_{\left[ {bu,t} \right]}},{Y_{\left[ {u,t} \right]}}} \right\}, \hfill \\
      b \in {\mathcal B},u \in {\mathcal U},t \in {\mathcal T}. \hfill \\
    \end{gathered}  \right.} \right\}.
\end{equation}
\par
Each BS chooses one available downlink power level from the downlink power space $\mathcal P$ to serve its associated user. Overall, the action space based on $\mathcal P$ is denoted by
\begin{equation}\label{AFBRL_Action_Space}
    {\mathcal A} = \left\{ {{A_t}\left| \begin{gathered}
      {A_t} = \left\{ {{A_{\left[ {bu,t} \right]}}} \right\},{A_{\left[ {bu,t} \right]}} \in {\mathcal P}; \hfill \\
      b \in {\mathcal B},u \in {\mathcal U},t \in {\mathcal T}. \hfill \\
    \end{gathered}  \right.} \right\}.
\end{equation}
\par
Power management not only results in a downlink power decline, but also causes an uncertain downlink data rate variation. The energy efficiency serves as the immediate reward in the reinforcement learning optimization process. In addition to a maximum immediate reward at the current time-step, we also optimize the long-term reward in terms of a cumulative energy efficiency. To this end, the long-term reward function \cite{Book_RL} is defined as the total discounted rewards from time-step $t$, and is computed as follows
\begin{equation}\label{AFBRL_Cumulative_Reward}
    \begin{aligned}%{gathered}
      {G_{\left[ {bu,t} \right]}} &= \sum\limits_{k = 0}^\infty  {{\lambda ^k}{R_{\left[ {bu,t + k + 1} \right]}}}  \hfill \\
       &= {R_{\left[ {bu,t + 1} \right]}} + \lambda {R_{\left[ {bu,t + 2} \right]}} + {\lambda ^2}{R_{\left[ {bu,t + 3} \right]}} +  \cdots  \hfill \\
       &= {R_{\left[ {bu,t + 1} \right]}} + \lambda \left( {{R_{\left[ {bu,t + 2} \right]}} + {\lambda }{R_{\left[ {bu,t + 3} \right]}} +  \cdots } \right) \hfill \\
       &= {R_{\left[ {bu,t + 1} \right]}} + \lambda {G_{\left[ {bu,t + 1} \right]}}. \hfill \\
    \end{aligned}%{gathered}
\end{equation}
where $\lambda \left( {0 < \lambda  \leq 1} \right)$ is the discount factor.
\par
Furthermore, the long-term reward is used to build the state-action-value function, which estimates how good it is to be at state $s$ with action $a$ according to policy $\pi$. Hence, the state-action-value function is given by
\begin{equation}\label{AFBRL_State_Value_Func}
    \!\!\!\begin{gathered}
      Q_{\left[ {bu,t} \right]}^\pi \left( {s,a} \right) \hfill \\
      \doteq {{\mathbb E}_\pi }\left[ {{G_{\left[ {bu,t} \right]}}\left| \begin{gathered}
      {S_{\left[ {bu,t} \right]}} = s,   {A_{\left[ {bu,t} \right]}} = a \hfill \\
        \end{gathered}  \right.} \right] \hfill \\
      = {{\mathbb E}_\pi }\left[ {{R_{\left[ {bu,t + 1} \right]}} + \lambda {G_{\left[ {bu,t + 1} \right]}}\left| \begin{gathered}
      {S_{\left[ {bu,t} \right]}} = s,  {A_{\left[ {bu,t} \right]}} = a \hfill \\
        \end{gathered}  \right.} \right] \hfill \\
      =\!\!\!\sum\limits_{s^{'}}\!{{P}\left( s^{'}|{s,a} \right)\!\!\left[ {{{R_{\left[ {bu,t + 1} \right]}}} + \lambda  \mathop {\max }\limits_{{a^{'}}} Q_{\left[ {bu,t + 1} \right]}^\pi \left( {{s^{'}},{a^{'}}} \right)}\!\right]} . \hfill \\
    \end{gathered}
\end{equation}
The expression in (\ref{AFBRL_State_Value_Func}) is the so-called Bellman equation for state $s$, and shows a relationship between the value of state $s$ and the values of its successive states.
\begin{figure*}[!t]
  \centering
  \includegraphics[width=5.5in]{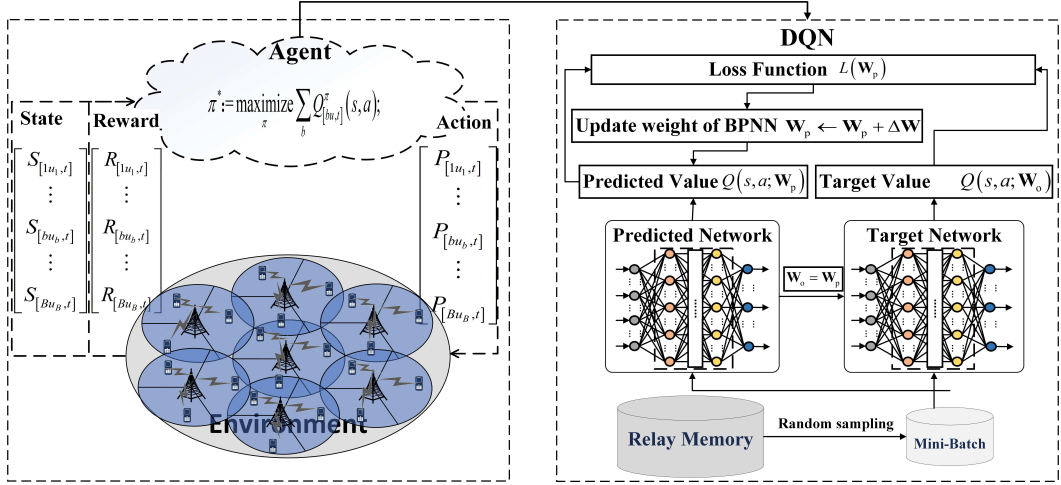}\\
  \caption{Workflow diagram of precise downlink power management Based on DQN.}\label{Fig_DQN_Workflow}
\end{figure*}
\par
The transition probability ${P}\left( s^{'}|{s,a} \right)$ depends on the optimization policy $\pi$ and the historical action selection in replay memory $\mathcal D$ defined in DQN. Therefore, ${\pi}\left( {s,a} \right)$, representing the transition probability at state $s$ with action $a$ according to the policy $\pi$, is calculated as follows
\begin{equation}\label{AFBRL_stateTransitionPro}
    \pi \left( {s,a} \right) = \left\{ \begin{gathered}
      \left( {1 - \varepsilon } \right) \cdot \frac{{\left| {{{\mathcal D}^a}} \right|}}{{\left| {\mathcal D} \right|}}, \qquad {1 - \varepsilon } ; \hfill \\
      \varepsilon  \cdot \frac{1}{{\left| {\mathcal P} \right|}}, \qquad \qquad \quad \varepsilon;  \hfill \\
    \end{gathered}  \right.
\end{equation}
where ${\left| {\mathcal D} \right|}$ is the size of replay memory $\mathcal D$. ${{\mathcal D} ^{a}}$ is the sample space related to action $a$ that is the subset of $\mathcal D$, and its size is ${\left| {\mathcal D}^a \right|}$. ${\left| {\mathcal P} \right|}$ is the size of the downlink power space $\mathcal P$.
\par
In MDP, policy $\pi$ specifies the action for each state. To achieve a good trade-off between exploration and exploitation, we adopts the following $\varepsilon$-greedy policy
\begin{equation}\label{AFBRL_greedy}
    {A_{\left[ {bu,t} \right]}} = \left\{ \begin{gathered}
      \mathop {\operatorname{argmax} }\limits_{{a^{'}}} Q\left( {{S_{\left[ {bu,t} \right]}},{a^{'}}} \right), \quad 1 - \varepsilon ; \hfill \\
      \operatorname{random} (A), \qquad \qquad \qquad \varepsilon . \hfill \\
    \end{gathered}  \right.
\end{equation}
\par
Hence, the long-term energy efficiency maximization in (\ref{SysMdl_ProblemFormation}) is equivalent to finding the optimal policy $\pi ^{*}$ such that the energy efficiency over time is maximized based on multiple constraints. Mathematically, the optimization problem in (\ref{SysMdl_ProblemFormation}) can be transformed into the following deep reinforcement learning problem
\begin{equation}\label{AFBRL_MaximizingReward}
    \begin{gathered}
    {\pi ^ * } := \mathop {\arg \max }\limits_{\pi} \sum\limits_b {Q_{\left[ {bu,t} \right]}^\pi \left( {s,a} \right)} ; \hfill \\
    s.t. \quad \sum\limits_{{k =0}}^{{T_{[bu,t]}}-1} {C_{[mn,t + k]}} = {V_{[bu,t]}}; \hfill \\
        \qquad \ \ \sum\limits_{b=1}^{B} {{\Delta C _{\left[ {bu,t} \right]}}}  >= 0 ; \hfill \\
        \qquad \ \ {P_{[b,t]}} \in {\mathcal P}; \hfill \\
        \qquad \ \ b \in {\mathcal B}, \ %\hfill \\
        u \in {\mathcal U}, \
        t \in {\mathcal T}. \hfill \\
    \end{gathered}
\end{equation}
where the optimization objective function, that is expressed by the cumulative energy efficiency in (\ref{SysMdl_ProblemFormation}), is replaced by finding the optimal policy based on the long-term reward in (\ref{AFBRL_State_Value_Func}).
\subsection{Power Management Based on Deep-Q-Network}\label{section:AFBRL_DQNA}
DQN improves Q-Learning with deep neural networks to make reinforcement learning solve complex and nonlinear optimization problems. The typical DQN framework is composed of the replay memory, mini-batch sample, target DQN, predicted DQN, weight updating, and loss function. By refining \textbf{Fig. }\ref{Fig_01_RRU_TU_Dis} with the RL framework, the workflow diagram of precise downlink power management based on DQN is illustrated in \textbf{Fig. }\ref{Fig_DQN_Workflow}. In DQN, the experience in each time-step is stored into the replay memory $\mathcal D$ for being accessed to update the DQN parameter. For example, after BS $b$ executes action ${A_{\left[ {bu,t} \right]}}$ at state ${S_{\left[ {bu,t} \right]}}$, it returns reward ${R_{\left[ {bu,t+1} \right]}}$ and transfers to the next state ${S_{\left[ {bu,t+1} \right]}}$. Hence, the tuple $\left( {{S_{\left[ {bu,t} \right]}},{A_{\left[ {bu,t} \right]}},{R_{\left[ {bu,t+1} \right]}},{S_{\left[ {bu,t + 1} \right]}}} \right)$ is stored into the replay memory $\mathcal D$ as follows
\begin{equation}\label{AFBRL_Replay_Memory}
    {\mathcal D} \leftarrow {\mathcal D} \cup \left\{ {\left( {{S_{\left[ {bu,t} \right]}},{A_{\left[ {bu,t} \right]}},{R_{\left[ {bu,t+1} \right]}},{S_{\left[ {bu,t + 1} \right]}}} \right)} \right\}.
\end{equation}
At each update process, the DQN parameter is updated based on the mini-batch samples memory ${\mathcal D}_m$ that comes from $\mathcal D$. The predicted network inputs the current state-action pair $\left( {{S_{\left[ {bu,t} \right]}},{A_{\left[ {bu,t} \right]}}} \right)$ and outputs the predicted value, i.e., ${Q}\left( {{S_{_{[bu,t]}}},{A_{_{[bu,t]}}};{{\mathbf{W}}_{\text{p}}}} \right)$.
The target network inputs the next state $s^{'}$ and outputs the maximum Q-value of the next state-action pair. Therefore, the target value of $\left( {{S_{\left[ {bu,t} \right]}},{A_{\left[ {bu,t} \right]}}} \right)$ is given by
\begin{small}
\begin{equation}\label{AFBRL_Target_Qvalue}
    \!\!\!\!\!\begin{gathered}
      Q\left( {{S_{\left[ {bu,t} \right]}},{A_{\left[ {bu,t} \right]}};{{\mathbf{W}}_{\text{o}}}} \right)% \hfill \\
       ={R_{\left[ {bu,t + 1} \right]}} + \lambda \mathop {\max }\limits_{{a^{'}}} Q\left( {{s^{'}},{a^{'}};{{\mathbf{W}}_{\text{o}}}} \right). \hfill \\
    \end{gathered}
\end{equation}
\end{small}
where $a^{'} ( a^{'} \in {\mathcal P} )$ is the candidate action for the next state.
\par
The loss function is used to measure whether the DQN parameter is optimized or not. In particular, the DQN parameter is optimized when the loss function is stable and tends to 0; and it is not optimized otherwise. Hence, based on the mini-batch sample set ${\mathcal D} _m$, the loss function measures how good is the predicted DQN as follows
\begin{equation}\label{AFBRL_Loss_Function}
    \begin{gathered}
      \operatorname{L}\left( {{\mathbf W} _{\text{p}}}, {{\mathbf W} _{\text{o}}} \right) \hfill \\
      = \frac{1}{{2\left| {{\mathcal D} _m} \right|}}\sum\limits_{k = 1}^{\left| {{\mathcal D} _m} \right|} {{{\left[ {Q\left( {{s_k},{a_k};{{\mathbf W} _{\text{p}}}} \right) - Q\left( {{s_k},{a_k};{{\mathbf W} _{\text{o}}}} \right)} \right]}^2}}, \hfill \\
    \end{gathered}
\end{equation}
where $\left| {{\mathcal D} _m} \right|$ is the size of mini-batch sample memory.
\begin{algorithm}[H]
\caption{Downlink Power Management using DQN}\label{alg:Power-Optimization-DQN}
\begin{algorithmic}
\STATE
\STATE {\bf Initialization: } $\mathcal B$, $\mathcal U$, $\mathcal P$, $\mathcal T$, $\mathcal D$, ${\mathcal D} _m$; %$\lambda$, $\varepsilon$, $T$, $N$.
        \FOR {$t=1$ to $T$ ($T$ is the number of episodes)}
            \STATE Obtain state {\small ${S_t} = \left\{ {{V_{\left[ {bu,t} \right]}},{X_{\left[ {bu,t} \right]}}} \right\},  b \in {\mathcal B}$} and $\zeta _t=0$;
            \FOR{$n=1$ to $N$ ($N$ is the maximum iteration)}
                \STATE Initialization vectors: ${\mathcal A} _n=\emptyset$ and ${\mathcal Q} _n^{\pi}=\emptyset$;
                \FOR {$b=1$ to $B$}
                    \IF {$1=={\varphi  _{[b,t]}}$}
                        \STATE Select action:
                        {\small ${A_{\left[ {bu,t} \right]}} = {P_{\left[ {b,t} \right]}} \sim \pi \left( {{S_{\left[ {bu,t} \right]}}} \right)$};
                        \STATE Update action vector: ${\mathcal A} _n\leftarrow {\mathcal A} _n \cup {A_{\left[ {bu,t} \right]}}$;
                    \ENDIF
                \ENDFOR
                \FOR {$b=1$ to $B$}
                    \STATE Calculate ${X_{[bu,t]}}$, $\Delta {C_{[bu,t]}}$, $R _{[bu,t]}$;
                    \STATE Update traffic volume: %\\ \qquad
                    {\small ${V_{[bu,t+1]}} \leftarrow {V_{[bu,t]}} - {C_{[bu,t]}}$};
                    \STATE Obtain the reward ${{R_{[bu,t+1]}}}$;
                    \STATE Transfer to state: %\\ \qquad
                    {\small ${S_{\left[ {bu,t + 1} \right]}} = \left\{ {{V_{\left[ {bu,t + 1} \right]}},{X_{\left[ {bu,t + 1} \right]}}} \right\}$};
                    \STATE Calculate $ Q_{\left[ {bu,t} \right]}^\pi$ based on (\ref{AFBRL_State_Value_Func}); %and
                    \STATE Update the value vector: ${\mathcal Q} _n^{\pi} \leftarrow {\mathcal Q} _n^{\pi} \cup {Q_{\left[ {bu,t} \right]}^{\pi}}$;
                \ENDFOR
                \IF {$\sum\limits_{b = 1}^B {\Delta {C_{[ {bu,t} ]}}}  >= 0$}
                    \STATE $\zeta _t$=1;
                \ENDIF
            \ENDFOR
            \STATE Find the optimal tuple based on ${\pi^*}: = \mathop {\operatorname{argmax} }\limits_{\pi} {\mathcal Q} _n^{\pi}$;
            \IF {$\zeta _t==1$}
                    \STATE Update $\mathcal D$ and record $n_t^{*}$ based on the optimal tuple;
            \ENDIF
            \IF {$\left( {\left| {\mathcal D} \right| > {\left| {{\mathcal D} _m} \right|}} \ \& \ 0==\!\!\!\!\!\mod (t,T_m)\right)$}
                \STATE Randomly extract mini-batch samples ${{\mathcal D} _m}$ from ${\mathcal D}$;
                \FOR {$k=1$ to ${\left| {{\mathcal D} _m} \right|}$}
                    \STATE Obtain $Q\left( {{s_k},{a_k};{{\mathbf W} _{\text{P}}}} \right)$ and $Q\left( {{s_k},{a_k};{{\mathbf W} _{\text{o}}}} \right)$;
                \ENDFOR
                \STATE Calculate $\operatorname{L}\left({\mathbf W} _{\text{P}} \right)$and update ${\mathbf W} _{\text{P}}$;
                \STATE After a fixed interval, update ${\mathbf W} _{\text{o}}$ as ${\mathbf W} _{\text{o}} = {\mathbf W} _{\text{P}}$.
            \ENDIF
        \ENDFOR
    \end{algorithmic}
\end{algorithm}
\par
The structure of precise downlink power management based on DQN is described in {\bf Algorithm \ref{alg:Power-Optimization-DQN}}, which includes two modules, named downlink power management optimization and the DQN training process. In the downlink power management optimization module, the optimal policy for each BS is learned to achieve a satisfactory reward considering downlink data rate variations described from line 3 to 22. During each optimization iteration, the vectors for storing the selected actions and their respective state-action-values of BSs are initialized in line 4. Each BS selects an action based on the $\varepsilon-$greedy policy in line 7, and stores it into the action vector ${\mathcal A} _n$ in line 8. From lines 11 and 18, a series of computations associated with the reward and state transition for each BS are performed. The satisfactory conditions for the training iteration in terms of the cumulative downlink data rate variation are defined from lines 19 to 21. Based on the state-action-value, the optimal policy is identified to obtain its corresponding tuple in line 23. The second module is described from line 27 to 34, wherein the DQN is trained by the transition pairs stored in replay memory $\mathcal D$. In line 28, the replay memory $\mathcal D$ and records $n_t^{*}$ are updated based on the optimal tuple. From lines 28 to 31, the mini-batch samples ${{\mathcal D} _m}$ are randomly extracted from $\mathcal D$ to calculate the predicted value and target value. In line 28, the loss function is calculated to update ${\mathbf W} _{\text p}$ based on (\ref{AFBRL_Loss_Function}), where $Q\left( {{s_k},{a_k};{{\mathbf W} _{\text{P}}}} \right)$ and $Q\left( {{s_k},{a_k};{{\mathbf W} _{\text{o}}}} \right)$ are, respectively, the predicted and target values of the $k^{th}$ mini-batch sample from ${{\mathcal D} _m}$. The gradient descent method is adopted to update ${{\mathbf W} _{\text{P}}}$ of the predicted network, and ${{\mathbf W} _{\text{o}}}$ is updated after a fixed interval in line 33.
\section{NUMERICAL RESULTS}\label{section:NumericalResults}
In this section, simulation results are presented to evaluate the performance of the proposed algorithm. We adopt the sleep scheme in \cite{ RTao2019TWC} as the benchmark, and evaluate performance improvement of the precise power management strategies that include the proposed algorithm and Q-learning. In particular, we first show the superiority of the proposed algorithm in terms of network energy efficiency over the Q-learning and sleep scheme. Next, the downlink average throughput and power of all BSs are analyzed based on the network energy efficiency definition in (\ref{SysMdl_CM_PowerEfficiency}). Furthermore, we present the effect of precise power management on the RSRP and interference reduction, which determines the downlink network throughput variation of all the BSs. Finally, the computation complexities of the proposed algorithm and Q-learning are verified by the optimization success ratio and iterations.
\begin{figure*}[!t]
  \centering
  \includegraphics[width=5.5in]{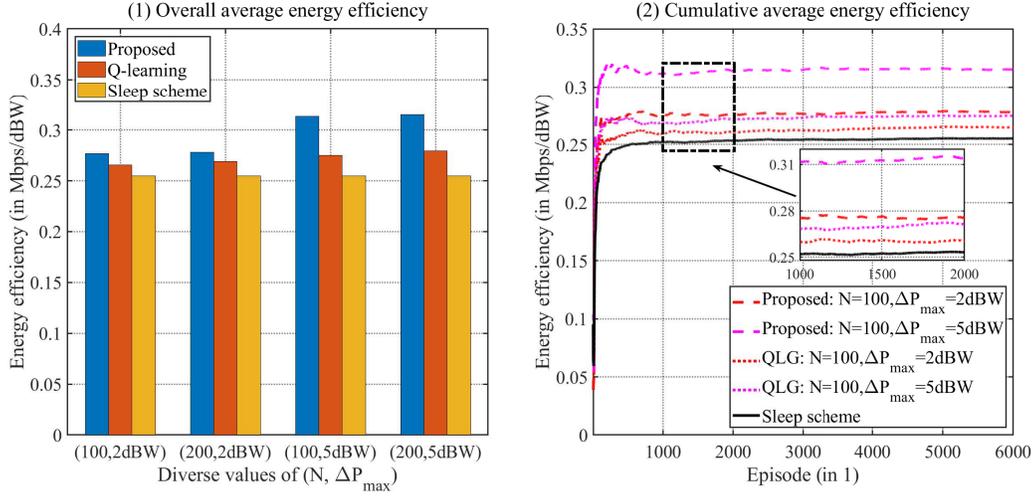}\\
  \caption{Cumulative average energy efficiency statistics and overall average energy efficiency statistics.}\label{Pic01_Reward_1}
\end{figure*}
\par
The basic simulation parameters are classified into the network topology parameters, BS equipment parameters, and algorithm setting parameters. The network topology parameters for the dense-RAN are the number of BSs $B=19$ and the maximum number of users scheduled at a time-step $U=57$. Meanwhile, hundreds and thousands of users are served by BSs in the multiple time-steps. The BS equipment parameters are $P_{\text{max}}=$15.2dBW, $w_0=$-125dBW, and $W=$10MHz; and the algorithm setting parameters are $T=20000$, $T_m=500$,
the reduced downlink power $\Delta P_{\max }=$2dBW or 5dBW,
$\lambda=0.9$, $\varepsilon=0.1$, $N=$100 or 200, $\left| {\mathcal D} \right|=5000$, and $\left| {\mathcal D}_m \right|=1000$.
\subsection{Energy Efficiency Analysis}\label{section:ConsumptionFactor}
In this subsection, we analyze the reward statistics of all algorithms based on overall and cumulative average network energy efficiency, which are given by
\begin{align}
\overline R _T^{\Sigma}  = \frac{1}{T}\sum\limits_{t = 0}^{T - 1} {\left( {\frac{1}{B}\sum\limits_{b = 1}^B {{R_{\left[ {bu,t} \right]}}} } \right)}, \label{NumbericalAnalyse_overallAveEE} \\
{\overline R _t^{\Sigma}} = \frac{1}{t+1}\sum\limits_{{t^{'}} = 0}^t {\left( {\frac{1}{B}\sum\limits_{b = 1}^B {{R_{\left[ {bu,{t^{'}}} \right]}}} } \right)}. \label{NumbericalAnalyse_cumulativeAveEE}
\end{align}
\par
The overall average network energy efficiency for the proposed algorithm, Q-learning, and sleep scheme are depicted in \textbf{Fig. }\ref{Pic01_Reward_1}-(1). First, the proposed algorithm achieves the largest overall average energy efficiency for different values of the tuple $(N, \Delta P _{\max})$ among all the algorithms. Second, when $\Delta P _{max}$ increases, the overall average energy efficiency of the proposed algorithm achieves better performance, while that of Q-learning improves slightly. Third, the iteration has little effect on the overall average energy efficiency of both the proposed algorithm and Q-learning.
Based on the overall average energy efficiency analysis, we set $(N, \Delta P _{\text{max}})=$(100, 2dBW) and $(N, \Delta P _{\text{max}})=$(100, 5dBW) to show the cumulative average energy efficiency of all the algorithms in \textbf{Fig. }\ref{Pic01_Reward_1}-(2). It can be learned that the proposed algorithm outperforms Q-learning based on the cumulative average energy efficiency, while Q-learning remains superior to the sleep scheme. Moreover, we enlarge the curves as shown in the small sub-image of \textbf{Fig. }\ref{Pic01_Reward_1}-(2), which shows that the energy efficiency fluctuates slightly over time due to a variable arrival rate. In particular, the fluctuating curve of the sleep scheme demonstrates that the variable request arrival rate leads to diversified BSs' sleep over the time-step. Next, we further analyze how the downlink power management affects the energy efficiency by controlling the downlink power and data rate of BSs in Section \ref{section:throughDownlinkTransmitPower}.
\subsection{Effect of Downlink Power Management}\label{section:throughDownlinkTransmitPower}
Because the downlink power and data rate of BSs are, respectively, the numerator and denominator of the energy efficiency, it is important to explore how the downlink power and data rate of BSs vary with the increasing number of episodes during the adaptive downlink power management process. Similar to ${\overline R} _t^{\Sigma}$, the cumulative average throughput of BSs is expressed as follows
\begin{equation}\label{NumbericalAnalyse_cumAveTgo}
    \overline C _t^\Sigma  = \frac{1}{{t + 1}}\sum\limits_{{t^{'}} = 0}^t {\left( {\frac{1}{B}\sum\limits_{b = 1}^B {{C_{\left[ {bu,{t^{'}}} \right]}}} } \right)} .
\end{equation}
The average downlink power of BSs in time-step $t$ is given by
\begin{equation}\label{NumbericalAnalyse_cumAveDownlinkPower}
    {\overline P _t} = 10 \cdot {\log _{10}}\left( {\frac{1}{{B}}\sum\limits_{b = 1}^B {{{10}^{\left( {{{{P_{\left[ {b,t} \right]}}} \mathord{\left/
     {\vphantom {{{P_{\left[ {b,t} \right]}}} {10}}} \right.
     \kern-\nulldelimiterspace} {10}}} \right)}}} } \right).
\end{equation}
Hence, the cumulative average power of BSs is given by
\begin{equation}\label{NumbericalAnalyse_CumAveTransPower}
      \overline P _t^\Sigma  = \frac{1}{{t + 1}}\sum\limits_{{t^{'}} = 0}^t {{{\overline P }_{{t^{'}}}}} . \hfill \\
\end{equation}
\begin{figure*}[!t]
\hrulefill
 \normalsize
 \begin{equation}\label{NumbericalAnalyse_DeltaRSRPITF}
    \left\{ \begin{gathered}
      \Delta {\overline P _t^S} = 10 \cdot {\log _{10}}\left( {\frac{1}{B}\sum\limits_{b = 1}^B {{\varphi _{\left[ {b,t} \right]}}\left( {{{10}^{^{\left( {{{{P_{\max }}} \mathord{\left/
     {\vphantom {{{P_{\max }}} {10}}} \right.
     \kern-\nulldelimiterspace} {10}}} \right)}}} - {{10}^{\left( {{{{P_{\left[ {b,t} \right]}}} \mathord{\left/
     {\vphantom {{{P_{\left[ {b,t} \right]}}} {10}}} \right.
     \kern-\nulldelimiterspace} {10}}} \right)}}} \right)} } \right), \hfill \\
      \Delta \overline P _t^I = 10 \cdot {\log _{10}}\left( {\frac{1}{B}\sum\limits_{b = 1}^B {\sum\limits_{{b^{'}} = 1,{b^{'}} \ne b}^B {{\varphi _{\left[ {{b^{'}},t} \right]}}\left( {{{10}^{^{\left( {{{{P_{\max }}} \mathord{\left/
     {\vphantom {{{P_{\max }}} {10}}} \right.
     \kern-\nulldelimiterspace} {10}}} \right)}}} - {{10}^{\left( {{{{P_{\left[ {{b^{'}},t} \right]}}} \mathord{\left/
     {\vphantom {{{P_{\left[ {{b^{'}},t} \right]}}} {10}}} \right.
     \kern-\nulldelimiterspace} {10}}} \right)}}} \right)} } } \right). \hfill \\
    \end{gathered}  \right. %\tag{31}
\end{equation}
\hrulefill
\end{figure*}
\begin{figure*}[!t]
  \centering
  \includegraphics[width=6.0in]{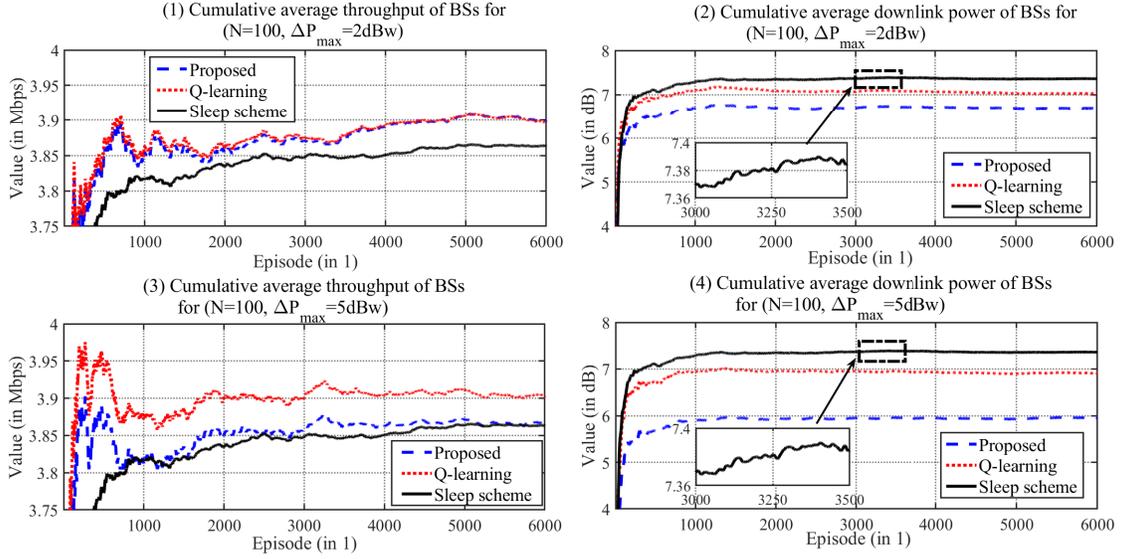}\\
  \caption{Statistics for the network throughput and downlink power with diverse $\Delta P _{max}$.}\label{Pic02_TgoPower_1}
\end{figure*}
\textbf{Fig. }\ref{Pic02_TgoPower_1} depicts the variations of the cumulative average downlink transmit power and data rate of BSs with increasing number of episodes for the settings $(N, \Delta P _{\text{max}})=$(100, 2dBW) and $(N, \Delta P _{\text{max}})=$(100, 5dBW). \textbf{Fig. }\ref{Pic02_TgoPower_1}-(1) and \textbf{Fig. }\ref{Pic02_TgoPower_1}-(3) imply that the average optimized downlink power of BSs has a slight effect on the average network throughput for the proposed algorithm and Q-learning in comparison to that of the sleep scheme, suggesting that it is feasible to optimize the downlink power without network throughput loss. \textbf{Fig. }\ref{Pic02_TgoPower_1}-(2) and \textbf{Fig. }\ref{Pic02_TgoPower_1}-(4) show the average downlink power of BSs for the proposed algorithm, Q-learning, and sleep scheme with increasing number of episodes. It can be learned that our proposed algorithm achieves better downlink power efficiency in comparison with Q-learning and sleep scheme, indicating that the downlink power optimization is almost the sole main contributor to the energy efficiency improvement. Meanwhile, we also enlarge the fluctuating sleep scheme curve as shown in the sub-images of \textbf{Fig. }\ref{Pic02_TgoPower_1}-(2) and -(4), indicating that the variable request arrival rate leads to diversified BSs' sleep over the time-step.
\begin{figure*}[!t]
  \centering
  \includegraphics[width=6.0in]{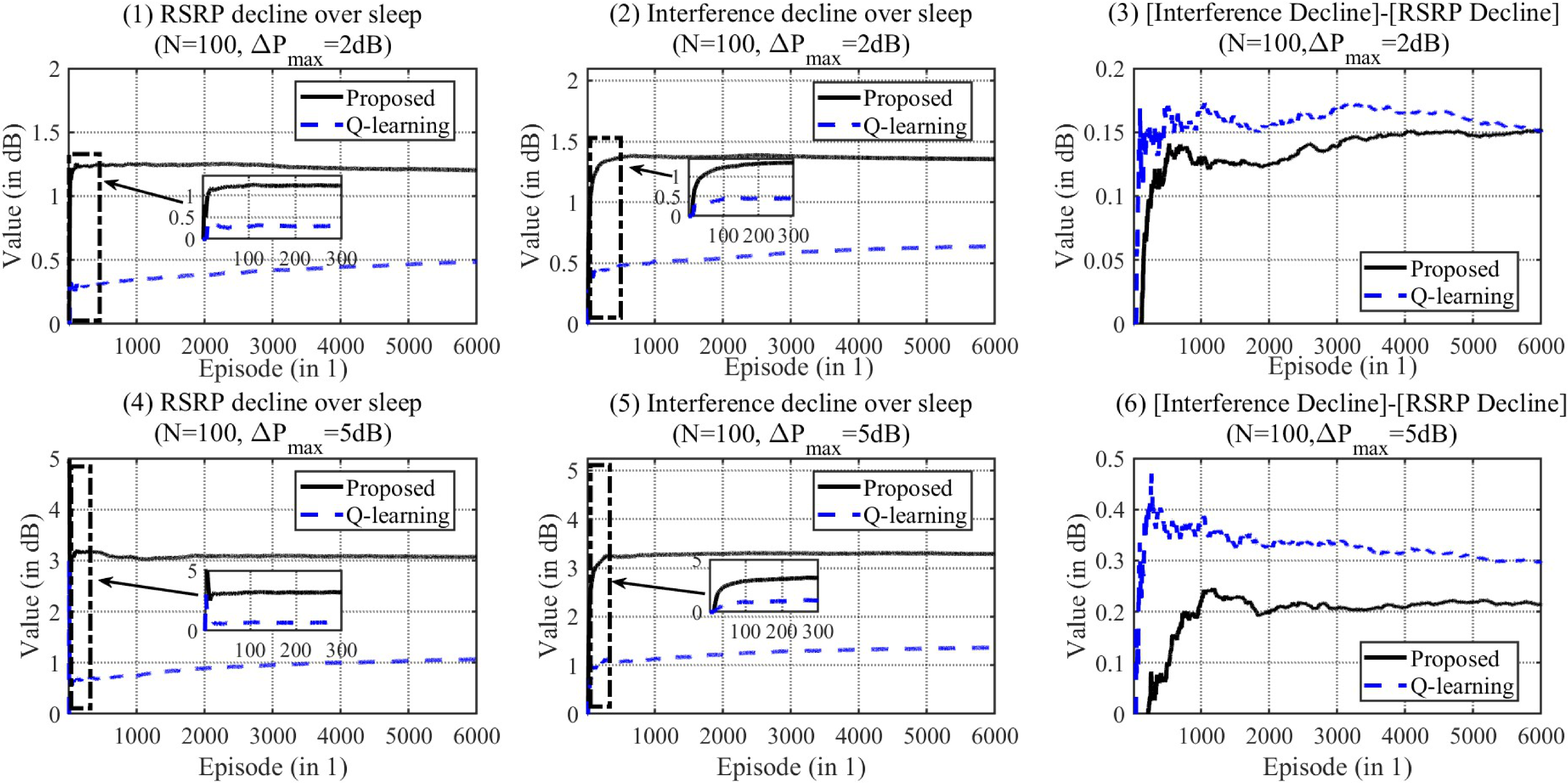}\\
  \caption{Statistics for the average RSRP and interference decline.}\label{Pic03_RSRPITF_1}
\end{figure*}
\par
The average downlink powers of BSs suffer from decline while the average throughputs of BSs hardly change, which inspires us to explore how downlink power management affects the network interference of BSs. In time step $t$, the average RSRP and interference decline of BSs in dBW are, respectively, expressed in (\ref{NumbericalAnalyse_DeltaRSRPITF}). Moreover, the cumulative average RSRP and interference decline of BSs are, respectively, given by
\begin{equation}\label{NumbericalAnalyse_CumAveRSRPITF}
   \left\{ \begin{gathered}
      \Delta \overline P _t^{S,\Sigma } = \frac{1}{{t + 1}}\sum\limits_{{t^{'}} = 0}^t {\Delta \overline P _{{t^{'}}}^S} , \hfill \\
      \Delta \overline P _t^{I,\Sigma } = \frac{1}{{t + 1}}\sum\limits_{{t^{'}} = 0}^t {\Delta \overline P _{{t^{'}}}^I} . \hfill \\
    \end{gathered}  \right.
\end{equation}
\textbf{Fig. }\ref{Pic03_RSRPITF_1} depicts declining variations of RSRP and interference for all BSs with increasing number of episodes for the settings $(N, \Delta P _{\text{max}})=$(100, 2dBW) and $(N, \Delta P _{\text{max}})=$(100, 5dBW). \textbf{Fig. }\ref{Pic03_RSRPITF_1}-(1) and \textbf{Fig. }\ref{Pic03_RSRPITF_1}-(4) show that the average RSRP decline of all the BSs of the proposed algorithm compared to the sleep scheme is much larger than that of Q-learning, whose trends match that of the average downlink power in \textbf{Fig. }\ref{Pic02_TgoPower_1}-(2) and \textbf{Fig. }\ref{Pic02_TgoPower_1}-(4), respectively. From \textbf{Fig. }\ref{Pic03_RSRPITF_1}-(2) and -(5), it can be learned that the average interference decline of all BSs of the proposed algorithm compared to the sleep scheme is also much larger than that of Q-learning. The enlarged curves of the sub-images in \textbf{Fig. }\ref{Pic03_RSRPITF_1}-(1), -(2), -(4) and -(5) show that the optimization process can converge in less 200 iterations, implying that the optimization is fast enough to cope with the necessary power adjustments. Moreover, the gaps between the average interference decline and the average RSRP decline for the settings $(N, \Delta P _{\text{max}})=$(100, 2dBW) and $(N, \Delta P _{\text{max}})=$(100, 5dBW) are, respectively, shown in \textbf{Fig. }\ref{Pic03_RSRPITF_1}-(3) and \textbf{Fig. }\ref{Pic03_RSRPITF_1}-(6). These figures indicate that better downlink power management of BSs can make the average interference decline larger than the average RSRP decline, which helps improve the average downlink networks throughput.
\begin{figure}[!t]
  \centering
  \includegraphics[width=3.0in]{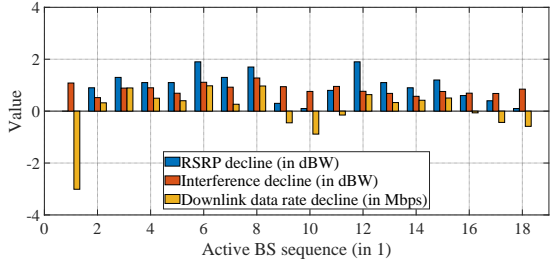}\\
  \caption{Decline of RSRP, interference, and data rate for active BSs.}\label{Pic05_RSRPITFTGO_1}
\end{figure}
However, the improvement of average downlink network throughput is less prominent due to the much smaller gap.
\par
For example, we randomly select one episode to analyze the RSRP decline, interference decline, and downlink data rate decline of active BSs for the setting $(N, \Delta P _{\text{max}})=$(100, 2dBW), which is illustrated in \textbf{Fig. }\ref{Pic05_RSRPITFTGO_1}. For any active BS, it can be learned that the downlink data rate increases when RSRP decline is smaller than interference decline, and that the downlink data rate declines when RSRP decline is larger than interference decline. However, the difference between RSRP decline and interference decline is a nonlinear function with the downlink data rate decline due to (\ref{SysMdl_CM_DR}). Based on (\ref{NumbericalAnalyse_DeltaRSRPITF}), the average RSRP decline and interference decline of all the BSs are calculated as 1.43dBW and 1.04dBW, respectively. The average downlink data rate decline of all BSs, denoted by $\frac{1}{B}\sum\limits_{b = 1}^B {\Delta {C_{\left[ {bu,t} \right]}}}$, decreases by 11.1Kb/s.
\par
\textbf{Fig.} \ref{Pic03_RSRPITF_1} and \ref{Pic05_RSRPITFTGO_1} show that power management of multi-BSs has different effect on declines of RSRP and interference for various BSs. The RSRP decline for one BS can simultaneously lead to the interference declines for its interference-related BSs, while the BS also enjoys the cumulative interference reduction from the RSRP decline of its interference-related BSs based on common sense of interaction. Moreover, \textbf{Fig.} \ref{Pic01_Reward_1} and \ref{Pic02_TgoPower_1} demonstrate that the optimization algorithm can converge quickly no matter how much BS reduces its downlink power, and that the proposed algorithm has good advantages. We can learn that collaborative multi-BSs power management for dense-RAN is to search the optimal power reduction combination of multi-BSs while considering the downlink data rate, and that DRL has advantages in handling the massive states' problem caused by multi-BSs through continuous learning. Therefore, it represents the joint optimization of performance and power for artificial intelligence (COPPAI) in dense-RAN.
\par
The above analysis implies that the adaptive downlink power management achieves the optimal downlink power for each BS. The optimal downlink power generates satisfactory RSRP decline and the corresponding interference decline at users, which weakens the loss of the average downlink network throughput. Hence, the optimal downlink power management improves the energy efficiency while guaranteeing the average downlink network throughput.
Moreover, the proposed algorithm achieves better communication performance in comparison with Q-learning and the sleep scheme. This is why the proposed algorithm can learn the dense-RAN state to more precisely manage the downlink power of BSs.
\subsection{Discussion on Optimization Complexity}\label{section:AnalysisIterationDRL}
In the proposed algorithm, the main time consuming step is to search for the optimal power value combination of all BSs, which requires $O(2^B)$ floating point operations (flops). To quantify the complexity of near-optimal solution, \textbf{Fig. }\ref{Pic01_Reward_1}-(1) shows that the two numbers of iterations, $N=100$ and $N=200$, have comparable effect on the energy efficiency improvement, which inspires us to further explore the computational complexity of both the proposed algorithm and Q-learning. In this subsection, the computational complexity of the proposed algorithm and Q-learning are analyzed in terms of the optimization success ratio and the number of iterations. First, the overall and cumulative average optimization success ratio are, respectively, defined as follows
\begin{equation}\label{NumbericalAnalyse_successStat}
    \left[ {\begin{array}{*{20}{c}}
      {\overline Z _T^{\Sigma }}&{\overline Z _t^{\Sigma }}
    \end{array}} \right] = \left[ {\begin{array}{*{20}{c}}
      {\frac{1}{T}\sum\limits_{t = 0}^{T - 1} {\zeta _t} }&{\frac{1}{{t + 1}}\sum\limits_{{t^{'}} = 0}^t {\zeta _t} }
    \end{array}} \right],
\end{equation}
where $\zeta_t=1$ represents optimization success and $\zeta_t =0$ is optimization failure.
Then the overall and cumulative average number of iterations are, respectively, defined as follows
\begin{equation}\label{NumbericalAnalyse_IterationStat}
   \!\!
   \left[ {\begin{array}{*{20}{c}}
      {\overline N _T^{*,\Sigma }}&{\overline N _t^{*,\Sigma }}
    \end{array}} \right] = \left[ {\begin{array}{*{20}{c}}
      {\frac{1}{T}\sum\limits_{t = 0}^{T - 1} {n_t^*} }&{\frac{1}{{t + 1}}\sum\limits_{{t^{'}} = 0}^t {n_t^*} }
    \end{array}} \right],
\end{equation}
where ${n_t^*}$ is the optimal number of iterations in time-step $t$.
\begin{figure*}[!t]
  \centering
  \includegraphics[width=6.0in]{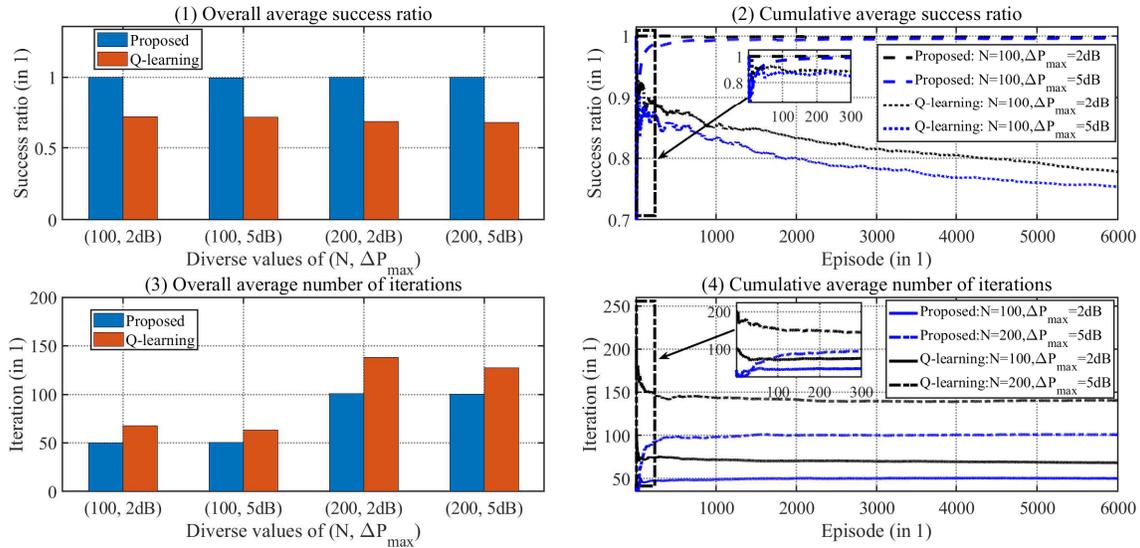}\\
  \caption{Statistics for the average optimization success ratio and iterations.}\label{Pic04_SuccessIteration_1}
\end{figure*}
\par
In \textbf{Fig. }\ref{Pic04_SuccessIteration_1}-(1) and \textbf{Fig. }\ref{Pic04_SuccessIteration_1}-(2), the overall and cumulative average optimization success ratio are, respectively, depicted with increasing number of episodes. In particular, \textbf{Fig. }\ref{Pic04_SuccessIteration_1}-(1) demonstrates that the overall average optimization success ratio of the proposed algorithm is approaching 1 and is much better than that of Q-learning. \textbf{Fig. }\ref{Pic04_SuccessIteration_1}-(2) shows that the cumulative average optimization success ratio of the proposed algorithm achieves rapid convergence in comparison with that of Q-learning. Meanwhile, the cumulative average optimization success ratio of the proposed algorithm remains stable with increasing number of episodes, while it drops slightly for Q-learning. Further, the curves in \textbf{Fig. }\ref{Pic04_SuccessIteration_1}-(2) are enlarged in the sub-image to demonstrate fast convergence for the necessary power adjustments. This is the reason that the Q-table of Q-learning fails to fully learn the dense-RAN state.
\par
In \textbf{Fig. }\ref{Pic04_SuccessIteration_1}-(3) and \textbf{Fig. }\ref{Pic04_SuccessIteration_1}-(4), the overall and cumulative average number of iterations are, respectively, depicted with increasing number of episodes. It can be learned that the overall and cumulative average number of iterations of the proposed algorithm are much smaller than those of Q-learning. Hence, like the communication performance, the computational complexity of the proposed algorithm is lower than Q-learning. It is insufficient for the Q-table of Q-learning to learn the dense-RAN state, which results in more explorations for Q-learning in comparison with the proposed algorithm.
\section{CONCLUSION}\label{section:conclusion}
In this paper, we propose a novel multi-BSs downlink power optimization algorithm to improve the long-term energy efficiency in dense-RAN, which is defined as the ratio of the downlink data rate by the downlink power. In the studied model, the transmitting traffic volume, generated from many users with unpredictable service demands, is integrated with the network throughput to formulate the constraints.
Based on the cloud-RAN operation scheme, we design a deep reinforcement learning framework to tackle the optimization problem.
Considering the Markov characteristics of traffic volume, we transform the maximization problem of long-term energy efficiency into the Bellman equation, and adopt a DQN to optimize multi-BSs downlink power management. In the studied DQN algorithm, BPNN is used to tune the DQN parameter to find the mapping approximation from the dense-RAN state to the cumulative reward. Simulation results show that the proposed algorithm demonstrates superiority of communication performance and computational complexity over Q-learning and the sleep scheme. Moreover, our numerical results demonstrate that precise downlink power management lowers the downlink power and guarantees the network throughput, which is key to improve the energy efficiency.

\section*{Acknowledgments}
The work of N. Al-Dhahir was supported by Erik Jonsson Distinguished Professorship at UT-Dallas

\bibliographystyle{IEEEtran}
\bibliography{bibtex}

\vfill

\end{document}